\def\gapprox{{_>\atop{^\sim}}}
\def\lapprox{{_<\atop{^\sim}}}

\def\cmmt{\rm {cm^{-2}}}
\def\cmsq{\rm {cm^2}}
\def\s-1{\rm {s^{-1}}}
\def\twco{$^{12}$CO}
\def\thco{$^{13}$CO}

\documentclass{aa}
\usepackage{graphics}
\begin{document}
\thesaurus{
03(11.05.2; 11.09.1 NGC 4194; 11.09.4; 11.19.3; 13.19.1; 13.19.3)}
%
%
%
\def\etal {et al.}
\def\ie {i.\,e.}
\def\etseq {{\em et seq.}}
\def\vs {{it vs.}}
\def\perse {{it per se}}
\def\adhoc {{\em ad hoc}}
\def\eg {e.\,g.}
\def\etc {etc.}
\def\ccpers {\hbox{${\rm cm}^3{\rm s}^{-1}$}}
\def\DEGR {\hbox{$^{\circ }$}}
\def\vlsr {\hbox{${v_{\rm LSR}}$}}
\def\vel {\hbox{${v_{\rm LSR}}$}}
\def\vhel {\hbox{${v_{\rm HEL}}$}}
\def\delv {\hbox{$\Delta v_{1/2}$}}
\def\dvel {\hbox{$\Delta v_{1/2}$}}
\def\TL {$T_{\rm L}$}
\def\TC {$T_{\rm c}$}
\def\TEX {$T_{\rm ex}$}
\def\TMB {$T_{\rm MB}$}
\def\TKIN {$T_{\rm kin}$}
\def\TREC {$T_{\rm rec}$}
\def\TSYS {$T_{\rm sys}$}
\def\TVIB {$T_{\rm vib}$}
\def\TROT {$T_{\rm rot}$}
\def\TDUST {$T_{\rm d}$}
\def\TASTAR {$T_{\rm A}^{*}$}
\def\TVIBST {$T_{\rm vib}^*$} 
\def\TB {$T_{\rm B}$}
\def \la{\mathrel{\mathchoice   {\vcenter{\offinterlineskip\halign{\hfil
$\displaystyle##$\hfil\cr<\cr\sim\cr}}}
{\vcenter{\offinterlineskip\halign{\hfil$\textstyle##$\hfil\cr
<\cr\sim\cr}}}
{\vcenter{\offinterlineskip\halign{\hfil$\scriptstyle##$\hfil\cr
<\cr\sim\cr}}}
{\vcenter{\offinterlineskip\halign{\hfil$\scriptscriptstyle##$\hfil\cr
<\cr\sim\cr}}}}}
\def \ga{\mathrel{\mathchoice   {\vcenter{\offinterlineskip\halign{\hfil
$\displaystyle##$\hfil\cr>\cr\sim\cr}}}
{\vcenter{\offinterlineskip\halign{\hfil$\textstyle##$\hfil\cr
>\cr\sim\cr}}}
{\vcenter{\offinterlineskip\halign{\hfil$\scriptstyle##$\hfil\cr
>\cr\sim\cr}}}
{\vcenter{\offinterlineskip\halign{\hfil$\scriptscriptstyle##$\hfil\cr
>\cr\sim\cr}}}}}
\def\RZWCO {${\cal R}_{2/1}^{\rm C^{18}O}$}
\def\RDRCO {${\cal R}_{3/2}^{\rm C^{18}O}$}
\def\RMSIO {${\cal R}_{5/2}^{\rm ^{28}SiO}$}
\def\RISOSIO {${ r }_{28/29}^{\rm SiO}$}
\def\RISOZSIO {${ r}_{29/30}^{\rm SiO}$}
\def\H0 {$H_{\rm o}$}
\def\mic {$\mu\hbox{m}$}
\def\micro {\mu\hbox{m}}
\def\SDOZ {\hbox{$S_{12\mu \rm m}$}}
\def\STWE {\hbox{$S_{25\mu \rm m}$}}
\def\SSIX {\hbox{$S_{60\mu \rm m}$}}
\def\SHUN {\hbox{$S_{100\mu \rm m}$}}
\def\solmass {\hbox{M$_{\odot}$}}
\def\solum {\hbox{L$_{\odot}$}}
\def\irlum {\hbox{$L_{\rm IR}$}}
\def\ohlum {\hbox{$L_{\rm OH}$}}
\def\blum {\hbox{$L_{\rm B}$}}
\def\numd {\hbox{$n\,({\rm H}_2$)}}                   
\def\rhounit {$\hbox{M}_\odot\,\hbox{pc}^{-3}$}
\def\kms {\hbox{${\rm km\,s}^{-1}$}}
\def\kmsyr {\hbox{${\rm km\,s}^{-1}\,{\rm yr}^{-1}$}}
\def\kmsmpc {\hbox{${\rm km\,s}^{-1}\,{\rm Mpc}^{-1}$}} 
\def\Kkms {\hbox{${\rm K\,km\,s}^{-1}$}}
\def\percc {$\hbox{{\rm cm}}^{-3}$}    
\def\cmsq  {$\hbox{{\rm cm}}^{-2}$}    
\def\cmsix  {$\hbox{{\rm cm}}^{-6}$}  
\def\arcsec {\hbox{$^{\prime\prime}$}}
\def\arcmin {\hbox{$^{\prime}$}}
\def\ffam {\hbox{$\,.\!\!^{\prime}$}}
\def\ffas {\hbox{$\,.\!\!^{\prime\prime}$}}
\def\ffM {\hbox{$\,.\!\!\!^{\rm M}$}}
\def\ffm {\hbox{$\,.\!\!\!^{\rm m}$}}
\def\ffs {\hbox{$\,.\!\!^{\rm s}$}}
\def\ffd {\hbox{$\,.\!\!^{\circ}$}}
\def\HI  {\hbox{HI}}
\def\HII {\hbox{HII}}
%
%
\def \AL {$\alpha $}    
\def \BE {$\beta $}     
\def \GA {$\gamma $}    
\def \DE {$\delta $}    
\def \EP {$\epsilon $}  
\def \alde {($\Delta \alpha ,\Delta \delta $)}
\def \MU {$\mu $}       
\def \TAU {$\tau $}     
\def \tapp {$\tau _{\rm app}$}
\def \tuns {$\tau _{\rm uns}$}
\def \RH {\hbox{$R_{\rm H}$}}         
\def \RT {\hbox{$R_{\rm \tau}$}}      
\def \BN  {\hbox{$b_{\rm n}$}}        
\def \BETAN {\hbox{$\beta _n$}}       
\def \TE {\hbox{$T_{\rm e}$}}         
\def \NE {\hbox{$N_{\rm e}$}}         
%
\def\MOLH {\hbox{${\rm H}_2$}}                    
\def\HDO {\hbox{${\rm HDO}$}}                     
\def\AMM {\hbox{${\rm NH}_{3}$}}                  
\def\NHTWD {\hbox{${\rm NH}_2{\rm D}$}}           
\def\CTWH {\hbox{${\rm C_{2}H}$}}                 
\def\TCO {\hbox{${\rm ^{12}CO}$}}                 
\def\CEIO {\hbox{${\rm C}^{18}{\rm O}$}}          
\def\CSEO {\hbox{${\rm C}^{17}{\rm O}$}}          
\def\CTHFOS {\hbox{${\rm C}^{34}{\rm S}$}}        
\def\THCO {\hbox{$^{13}{\rm CO}$}}                
\def\WAT {\hbox{${\rm H}_2{\rm O}$}}              
\def\WATEI {\hbox{${\rm H}_2^{18}{\rm O}$}}       
\def\CYAN {\hbox{${\rm HC}_3{\rm N}$}}            
\def\CYACFI {\hbox{${\rm HC}_5{\rm N}$}}          
\def\CYACSE {\hbox{${\rm HC}_7{\rm N}$}}          
\def\CYACNI {\hbox{${\rm HC}_9{\rm N}$}}          
\def\METH {\hbox{${\rm CH}_3{\rm OH}$}}           
\def\MECN {\hbox{${\rm CH}_3{\rm CN}$}}           
\def\METAC {\hbox{${\rm CH}_3{\rm C}_2{\rm H}$}}  
\def\CH3C2H {\hbox{${\rm CH}_3{\rm C}_2{\rm H}$}} 
\def\FORM {\hbox{${\rm H}_2{\rm CO}$}}            
\def\MEFORM {\hbox{${\rm HCOOCH}_3$}}             
\def\THFO {\hbox{${\rm H}_2{\rm CS}$}}            
\def\ETHAL {\hbox{${\rm C}_2{\rm H}_5{\rm OH}$}}  
\def\CHTHOD {\hbox{${\rm CH}_3{\rm OD}$}}         
\def\CHTDOH {\hbox{${\rm CH}_2{\rm DOH}$}}        
\def\CYCP {\hbox{${\rm C}_3{\rm H}_2$}}           
\def\CTHHD {\hbox{${\rm C}_3{\rm HD}$}}           
\def\HTCN {\hbox{${\rm H^{13}CN}$}}               
\def\HNTC {\hbox{${\rm HN^{13}C}$}}               
\def\HCOP {\hbox{${\rm HCO}^+$}}                  
\def\HTCOP {\hbox{${\rm H^{13}CO}^{+}$}}          
\def\NNHP {\hbox{${\rm N}_2{\rm H}^+$}}           
\def\CHTHP {\hbox{${\rm CH}_3^+$}}                
\def\CHP {\hbox{${\rm CH}^{+}$}}                  
\def\ETHCN {\hbox{${\rm C}_2{\rm H}_5{\rm CN}$}}  
\def\DCOP {\hbox{${\rm DCO}^+$}}                  
\def\HTHP {\hbox{${\rm H}_{3}^{+}$}}              
\def\HTWDP {\hbox{${\rm H}_{2}{\rm D}^{+}$}}      
\def\CHTWDP {\hbox{${\rm CH}_{2}{\rm D}^{+}$}}    
\def\CNCHPL {\hbox{${\rm CNCH}^{+}$}}             
\def\CNCNPL {\hbox{${\rm CNCN}^{+}$}}             
%
%
\def\In {\hbox{$I^{n}(x_{\rm k},y_{\rm k},u_{\rm l}$})}
\def\Iobs {\hbox{$I_{\rm obs}(x_{\rm k},y_{\rm k},u_{\rm l})$}}
\def\Ingl {I^{n}(x_{\rm k},y_{\rm k},u_{\rm l})}
\def\Iobsgl {I_{\rm obs}(x_{\rm k},y_{\rm k},u_{\rm l})}
\def\Pbgl {P_{\rm b}(x_{\rm k},y_{\rm k}|\zeta _{\rm i},\eta _{\rm j})}
\def\Pbgm {P(x_{\rm k},y_{\rm k}|r_{\rm i},u_{\rm l})}
\def\Pbgn {P(x,y|r,u)}
\def\Pugm {P_{\rm u}(u_{\rm l}|w_{\rm ij})}
\def\Pdem {P_{\rm b}(x,y|\zeta (r,\theta ),\eta (r,\theta ))} 
\def\Pden {P_{\rm u}(u,w(r,\theta ))}
\def\greekgl {(\zeta _{\rm i},\eta _{\rm j},u_{\rm l})}
\def\greekg1 {(\zeta _{\rm i},\eta _{\rm j})}
\title{Complex molecular gas structure in the Medusa merger}
\author{S.~Aalto\inst{1}, S.~H\"uttemeister\inst{2}}
\offprints{S. Aalto, susanne@oso.chalmers.se}
\institute{
 Onsala Rymdobservatorium, S - 43992 Onsala, Sweden
\and
Radioastronomisches Institut der Universit\"{a}t Bonn,
 Auf dem H\"{u}gel 71, D - 53121 Bonn, Germany
}
\date{Received  / Accepted  }
\titlerunning{Molecular Gas in the Medusa Merger}
\maketitle
\begin{abstract}

High resolution OVRO aperture synthesis maps of the
\twco\ 1--0 emission in the ``Medusa'' galaxy merger (NGC~4194) reveal
the molecular emission being surprisingly extended. The 
\twco\ emission is distributed on a total scale of 25$''$ (4.7 kpc)
--- despite the apparent advanced stage of the merger.
The complex, striking \twco\ morphology 
occupies mainly the center and the north-eastern part of the main optical body.
The extended \twco\ flux is tracing two prominent dust lanes:
one which is crossing the central region at right angle (with respect to the
optical major axis) and a second which curves to the north-east and then into
the beginning of the northern tidal tail. 

The bulk of the \twco\ emission (67\%) can be found in a complex starburst region
encompassing the central 2\,kpc. The molecular gas
is distributed in five major emission regions of typical size 300\,pc.
About 15\% of the total \twco\ flux is found in a bright region 
1$\hbox{$\,.\!\!^{\prime\prime}$}$5 south of the
radio continuum nucleus. 
We suggest that this region together with the kpc sized
central starburst is being fueled by gas flows along the central dust lane. 
We discuss the merger history of NGC~4194 and suggest that it may be the result of
a early-type/spiral merger with a shell emerging to the south of the center.

The total molecular mass in the system is estimated to be at most $2 \times 10^9$ M$_{\odot}$,
depending on which \twco\ - H$_2$ conversion factor is applicable. 
The high \twco/\thco\ 1--0 intensity ratio, $\approx 20$, indicates highly excited physical
conditions
in the interstellar medium showing that the starburst has a big impact on its surrounding
ISM. At the current rate of star formation, the gas will be consumed within 40 million years.

\end{abstract} 


\section{Introduction}

The Medusa merger, NGC~4194, (Table 1; Fig.\,1) belongs to a class of lower luminosity 
($L_{\rm FIR} = 8.5 \times 10^{10}$ L$_{\odot}$ at $D$=39 Mpc)
mergers, an order of magnitude fainter than well known Ultra Luminous Galaxies (ULIRGs
($L_{\rm FIR} \gapprox 10^{12}$ L$_{\odot}$))
such as Arp~220, but with an extended region of intense star formation (Prestwhich \etal\ 1994;
Armus \etal\ 1990). 
As the name indicates, the morphology of the Medusa merger
is as spectacular as that of the brighter objects. The ``hair'' or ``tentacles''
are a patchy tidal tail stretching out 60$''$ (11.3 kpc) north of the main body,
which appears twisted. 
There also is a sharp, curved feature (and a very faint tail) to the south.

NGC~4194 has been mapped in \twco\ 1-0 and 2-1 with a single dish telescope by Casoli \etal\ (1992) and
detected in the more highly excited \twco\ 3-2 line by Devereux \etal\ (1994). It
is one of the handful known mergers (e.g. Arp~220, IC~694, NGC~6240) with unusually
high \twco/\thco\ 1--0 intensity ratios, ${\cal R} >$20 (e.g. Aalto \etal\ 1991b; Casoli \etal\ 1992).
In contrast, the majority of large galaxies exhibit ${\cal R}$ in the range 5 --- 15
(e.g. Aalto \etal\ 1995). One feature these HR (High Ratio) mergers have in common is a high FIR
$f(60 \mu{\rm m})/f(100 \mu{\rm m})$ flux ratio $\gapprox 0.8$ -- which likely is an indication
of a high dust temperature (e.g. Aalto \etal\ 1991a, 1991b). High resolution \twco\ maps
of several luminous ($L_{\rm IR} \gapprox 10^{11}$ L$_{\odot}$) HR mergers reveal that nearly
all the molecular gas is concentrated within a few hundred parsecs of the center
(e.g. Scoville \etal\ 1989; Bryant \& Scoville 1996; Downes \& Solomon 1998). 

The resulting large central surface densities require high pressures in hydrostatic
equilibrium and the molecular ISM becomes highly turbulent. 
Large turbulent line widths in molecular clouds will result in reduced
optical depths in the \twco\ 1--0 line and the \twco/\thco\ 1--0 line ratio will be elevated compared
to the optically thick ratio (for a discussion see Aalto \etal\ 1995). So for the luminous HR mergers
this could be one important mechanism
behind the large observed \twco/\thco\ line ratios (Aalto \etal\ 1995), aside from the additional
effect from large kinetic temperatures. Will the lower
luminosity HR galaxy NGC~4194 also follow this pattern of a compact nuclear starburst, or will an
extended \twco\ distribution require a different dominant mechanism behind the elevated ${\cal R}$?

Most of the high resolution imaging effort of \twco\ in mergers has been directed toward the
ULIRGs  and not much
is known, statistically, of the more modest luminosity merging galaxies. 
Apart from addressing the question of the compactness of the \twco\ emission
in an HR merger, a \twco\ image of NGC~4194 will improve the statistics of the \twco\
distribution in moderate luminosity mergers.

The elevated values of ${\cal R}$ in hot dust mergers have also been interpreted as a global 
underabundance in \thco\ with respect to \twco\ and not an effect of reduced optical depths in
\twco\ (e.g. Casoli \etal\ 1992). However, the moderate optical depth ($\tau \approx 1$)
in \twco\ 1--0 towards centers of galaxies in general, and ULIRGs in particular, is now a
reasonably well established fact (e.g. Aalto \etal\ 1997; Dahmen \etal\ 1998; Downes \& Solomon 1998).

We have imaged NGC 4194 in the \twco\ 1--0 line with OVRO to study the distribution
and kinematics of the molecular gas in an HR merger galaxy of moderate 
luminosity. We have also obtained single dish spectra of the 1--0 transitions of
\twco, \thco\ and HCN. The observations are presented in Sect.\ 2; the results in
Sect.\ 3; the consequences of these findings are discussed in terms of merger history, 
central gas surface densities
and final fate of the molecular gas in Sect.\ 4.

\section{Observations}

\begin{table*}
\caption{\label{prop} Properties of NGC~4194$^{\rm a)}$}
\begin{tabular}{lrrrcccccc}
\multicolumn{1}{c}{R.A. (1950.0)} & 
\multicolumn{1}{c}{Dec (1950.0)} &
Type & $m_b$ & Size (blue) & PA & $L_{\rm FIR}$ & ${f(60\mu{\rm m}) \over f(100 \mu{\rm m})}$ & $M$(HI)
& $M_{\rm dyn}$\\
\hline \\
$12^{\rm h}\ 11^{\rm m}\ 41.22^{\rm s}$ & $54^{\circ}\ 48'\ 16''$$^{\rm b)}$ &
Pec$^{\rm c)}$ & 13.0$^{\rm c)}$ & 2\hbox{$\,.\!\!^{\prime}$}3
$\times$ 1\hbox{$\,.\!\!^{\prime}$}6 $^{\rm c)}$ & 
146$^{\rm d)}$ & $8.5 \cdot 10^{10}$\solum$^{\rm e)}$ & 0.97 &
 $2.2 \times 10^{9}\,$ M$_{\odot}^{\rm f)}$ & $7 \times 10^9$ M$_{\odot}^{\rm g)}$
\\
\hline \\
\end{tabular} \\
a): Other aliases: UGC~07241, Arp~160, Mrk~201, VV~261, I~Zw~033 \\
b): Radio continuum peak position (Condon et al.\ (1990) \\
c): from the UGC catalogue (Nilson 1973) \\
d): as measured the bright main body of the optical R-band image (from Mazzarella \& Boroson)\\
e): (10$\mu$m - 100$\mu$m), Deutsch \& Willner (1987) (for $D$=39 Mpc)\\
f): Thuan and Martin (1981) (for $D$=39 Mpc)\\
g): Adopting the inclination $i \approx 40^{\circ}$ and using Fig.\,4, we get 
$V_{\rm rot}$=115 \kms. The dynamical mass is 
$M_{\rm dyn} = 2.3 \times 10^8 ({V_{\rm rot} \over 100})^2 ({R \over 100})$ M$_{\odot}$ where
$V_{\rm rot}$ is in \kms\ and $R$ in pc. 
For the inner 8$''$ (1.5 kpc) radius the dynamical mass is 
$M_{\rm dyn}=7 \times 10^9$ M$_{\odot}$. In this unrelaxed system,
one should of course use dynamical masses, derived from the assumption of regular rotation, with care.

\end{table*}

\begin{figure}
\resizebox{\hsize}{!}{\includegraphics{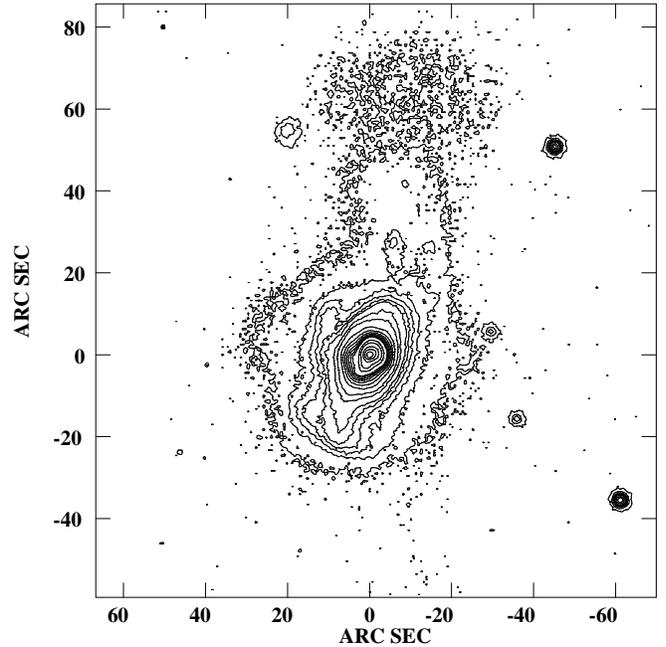}}
\caption{\label{something} Contour plot of an R-band CCD image of NGC~4194 
(Mazzarella \& Boroson 1993). The levels are logarithmic to clearly show
the faint structure, such as the northern tidal tail.  }
\end{figure}

We have obtained high resolution maps of \twco\ 1--0 using
the Caltech six-element Owens Valley Radio Observatory
(OVRO) millimeter array. Two tracks were taken in the low resolution configuration
in September and October 1996 and two in the high resolution configuration
in December 1996. The naturally weighted synthesized beam size is 
$2\hbox{$\,.\!\!^{\prime\prime}$}5 \times 2\hbox{$\,.\!\!^{\prime\prime}$}0$
($472 \times 378$ pc for $D$=39 Mpc) and the beam position angle (BPA) is --66$^{\circ}$.
The sensitivity of this map is 11 mJy beam$^{-1}$ channel$^{-1}$,
corresponding to 0.19 K channel$^{-1}$. For a beamsize 
of $2$\hbox{$\,.\!\!^{\prime\prime}$}3 a
brightness temperature ($T_{\rm B}$) of 1 K corresponds to 
57 mJy at $\lambda$=2.6 mm.

For the (naturally weighted) high resolution data
only, the beamsize is $2\hbox{$\,.\!\!^{\prime\prime}$}0 \times 
1\hbox{$\,.\!\!^{\prime\prime}$}7$ and BPA is --71$^{\circ}$.  For a resolution of 
$1$\hbox{$\,.\!\!^{\prime\prime}$}85 
a brightness temperature of 1 K corresponds to 36 mJy. 

System temperatures were 500--600 K. The quasar 1150+497 was 
used for phase calibration and Neptune and Uranus for
absolute flux calibration.  
The digital correlator, centered at $V_{\rm LSR} = 2560$ \kms, provided
a total velocity coverage of  1170 \kms.
Data were binned to 4 MHz resolution, corresponding to 
10 \kms, and to construct the map, the resolution was reduced to 20 \kms.


\begin{table}
\caption{\label{prop} Single dish results}
\begin{tabular}{lrr}
Line & $\theta$ ($''$) & $\int{T_{mb}}$ (K \kms) \\
\hline \\
\twco\ & 33 & $17 \pm 2$ \\
\thco\ & 34 & $0.9 \pm 0.2$$^{\rm a}$ \\
HCN & 43 & $\lapprox 0.4$$^{\rm b}$ \\ 
\hline \\
\end{tabular} \\
a): The \twco/\thco\ 1--0 intensity ratio is $19 \pm 4$. \\
b): The \twco/HCN 1--0 intensity ratio is $\gapprox 25$, corrected for
beamsize.
\end{table}

Single dish spectra of the 1--0 transition of \twco, \thco\ and HCN  were obtained toward
the center of NGC~4194 with the OSO 20m telescope. The observations were carried out
in October 1999, and the 
system temperatures were typically 400 K for \twco\ and \thco\ and 200--300 K for HCN.
The pointing accuracy was checked on SiO masers and was found to be about 3$''$.
Beam sizes and efficiencies are listed in Table 2, together with the results.

\begin{figure*}
\resizebox{16cm}{!}{\includegraphics{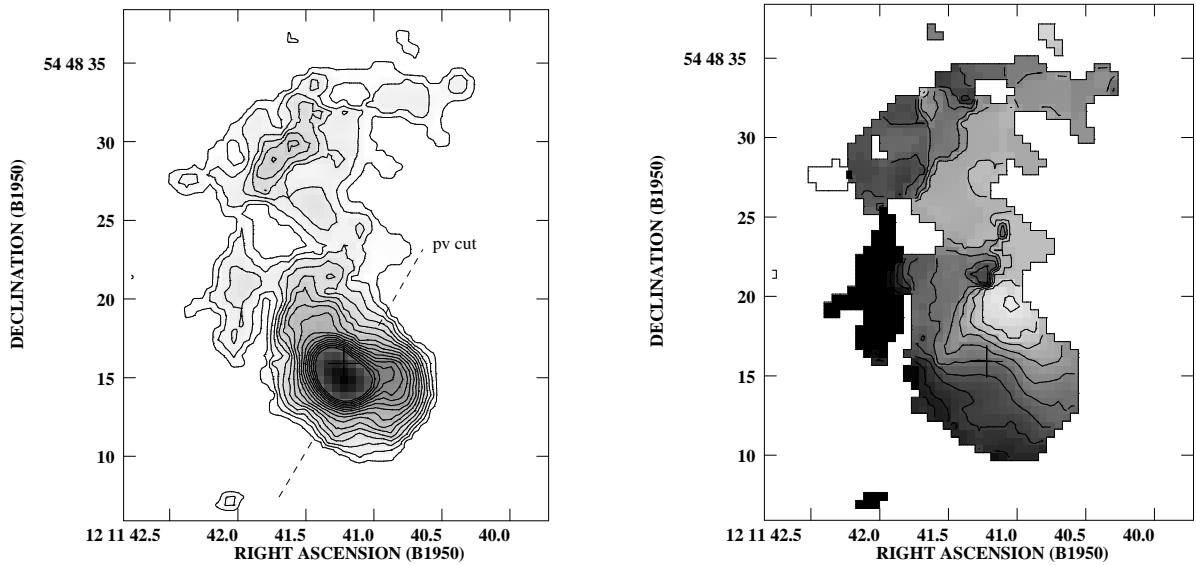}}
\caption{\label{something}  {\bf a)} Integrated intensity of the naturally weighted \twco\ map.
Contour range is (0.54, 1.6, 2.7, 3.8, ..., 31) Jy beam$^{-1}$ \kms. The grayscale 
ranges from 0.7 (light) to 27 (dark) Jy beam$^{-1}$ \kms. The radio continuum position
is marked with a cross. The total flux in map is 170 Jy \kms. The dashed line indicates
the location of the position velocity cut in Fig.\,6 {\bf b)}: The velocity field.
Contours range from 2400 to 2640 \kms, with spacing 24 \kms . 
The grayscale ranges from 2400 (light) to 2650 (dark) \kms}
\end{figure*}

\section{Results}
\subsection{Morpholology and kinematics of the molecular
distribution}

{\it Integrated intensity:} The naturally weighted \twco\ emission (Fig.\,2a) is 
distributed on a scale of $30''$ (5.7 kpc). The bulk of the
emission (67 \%) is concentrated in an 10$''$ (1.9 kpc) sized oval structure 
centered on the optically bright main body.  Fainter, patchy
emission curves out to the north from the brighter emission. 

An overlay of the \twco\ distribution on an R-band image (Fig.\,3) reveals that
the general \twco\ morphology is quite different from that of the main optical
body of the galaxy. The molecular gas extends from the center to the north-eastern part of
the object, and the {\it major axis of the central \twco\ intensity distribution is
almost perpendicular to that of the optical body}. 
The position angle (PA) of the
\twco\ oval intensity distribution of the bright emission is 
$\approx 49^{\circ}$ and the FWHM source size is 
$4\hbox{$\,.\!\!^{\prime\prime}$}2 \times 2$\hbox{$\,.\!\!^{\prime\prime}$}5 The position
angle of
the optical emission is 146$^{\circ}$.

The curved northern \twco\ emission reaches
into the base of the tidal tail, and some faint emission is also detected near a
bright patch (indicated with an arrow in Fig.\,3) in the tail itself. In Fig.\,3, the \twco\
emission appears to be inside of an optical protrusion, or arm, in the east (marked with
an arrow in Fig.\,3).
Armus \etal\ (1990) describe this feature as a ``spur'' or a ``jet''. The blue band
image in Arp's Catalog of Peculiar Galaxies (1966) gives instead a strong indication of a
foreground eastern dust lane blocking the background emission and creating
the impression of a protruding feature separate from the main body.
The molecular emission traces the space between the ``spur'' and the main
body which is consistent with the interpretation of a dust lane.

The total \twco\ flux in NGC~4194 detected by OVRO is 170 Jy \kms\ which, for a standard
\twco\ to H$_2$ mass conversion factor, translates to $2.3 \times 10^9$ M$_{\odot}$ ($0.92 \times 10^4 S
\Delta V D^2$ M$_{\odot}$ for $X$ = N(H$_2$)/I(\twco) = $2.3 \times 10^{20}$ $\cmmt$ (K \kms)$^{-1}$) of molecular
gas. Within calibrational error bars, we recover all of the single dish flux.
The peak flux is 27 Jy beam$^{-1}$ \kms\ (16\% of the total flux)
at $\alpha$: $12^{\rm h}\ 11^{\rm m} \ 41.22^{\rm s}$ ;
$\delta$: $54^{\circ}\ 48'\ 14\hbox{$\,.\!\!^{\prime\prime}$}4$.
which translates to a molecular mass of $3.8 \times 10^8$ M$_{\odot}$ within a radius of 240 pc.
However, we suspect that the conversion factor overestimates the molecular mass in this
system --- in particular in the
center (see Sect.\,4.3.1.).

\begin{figure}
\resizebox{6cm}{!}{\includegraphics{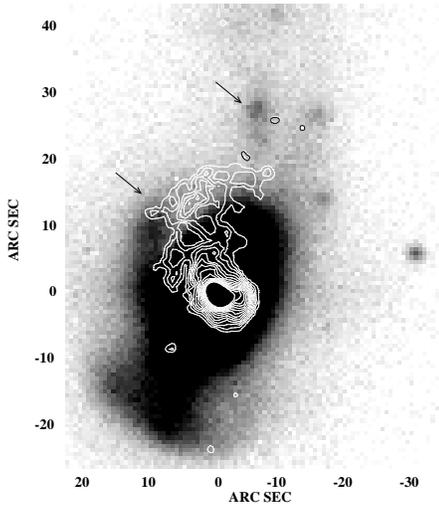}}
\caption{\label{something} Overlay of the \twco\ contours (white, apart from clouds in the
tail marked in black) on a greyscale, overexposed
optical R-band image (Mazzarella \& Boroson 1993). The left arrow
indicates the eastern expansion and the right one points to one of the bright
condensations in the northern tidal tail.
 }
\end{figure}

{\it The velocity field:} In Fig.\,2b we present the \twco\ velocity field
which shows the complex dynamics
of NGC~4194. Close to the center there seems to be organized
rotation of the gas (even though the velocity contours are far from a regular 
spider diagram) with a position angle of 160$^{\circ}$ --- close
to the optical major axis. 
Thus, the major axis of the rotation of the gas follows the PA of the optical light
distribution, rather than the distribution of the \twco\ emission.
A position-velocity (pV) cut at PA 160$^{\circ}$ (Fig.\,4) indicates a 
solid-body rotation curve which starts to flatten 
(or fall) at a radius of 2$''$ (380 pc). The Figure also shows that the \twco\ peak is
offset from the kinematic center by $1''$.
The projected rotational velocity is 90 \kms . Thus, for an inclination of 40$^{\circ}$
(from the optical isophotes) the deprojected rotational velocity is 140 \kms\ in the
inner 16$''$ (3 kpc). The systemic velocity at the kinematic center is 2510 \kms.

\begin{figure}
\resizebox{6cm}{!}{\includegraphics{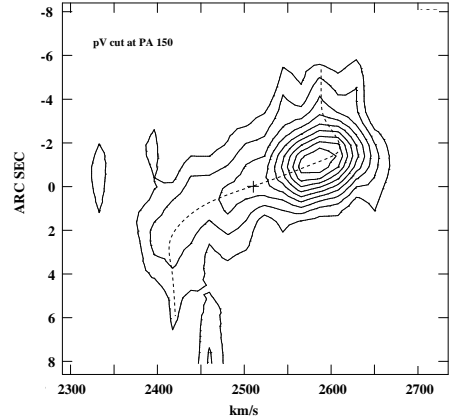}}
\caption{\label{something} pV cut at PA 160$^{\circ}$ across the
central \twco\ body. The location of the cut is indicated in Fig.\,2a.
The cross marks the estimated kinematic center of a solid body rotating
structure. }
\end{figure}

The extended northern gas in general does not move according to the pattern established
in the inner region. The eastern
part of the emission is at higher velocities than the western emission and velocity
discontinuities can be traced along the curved ridge at the transition between the
velocity systems. The highest velocity in the system, 2700 \kms, can be found on the
eastern dust lane at $\delta$: $54^{\circ}\ 48'\ 22''$.
The velocity of the gas in the eastern dust lane then gradually
drops along the lane, until it reaches the tidal tail where
the velocity of the gas clumps is 2580 \kms.

\begin{figure*}
\resizebox{14cm}{!}{\includegraphics{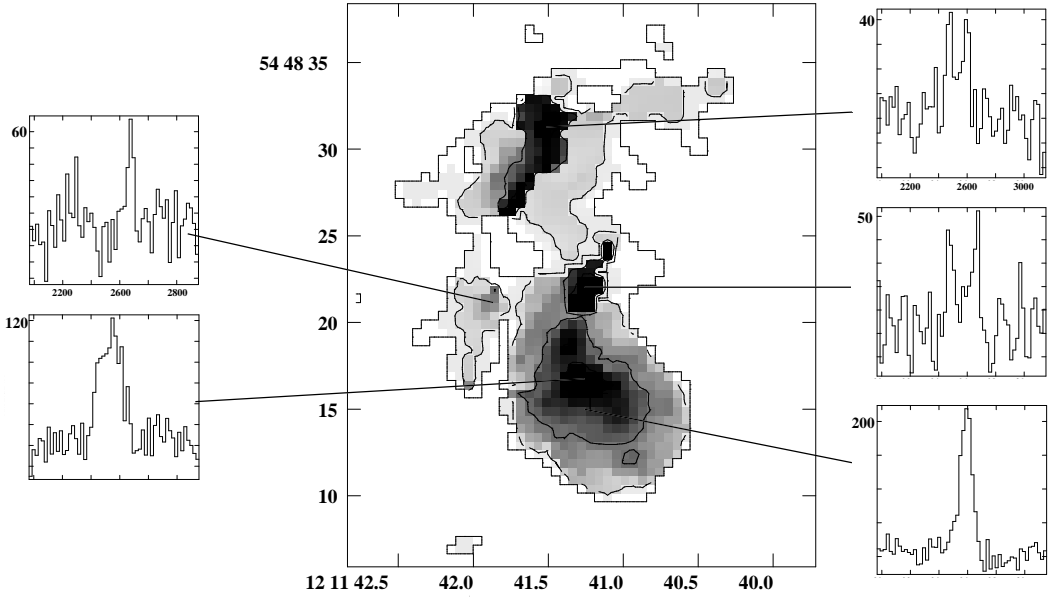}}
\caption{\label{something} Dispersion map with sample spectra. The
contours are (12, 36, 60)\,\kms\ and the grayscale ranges
from 6 (light) to 60 (dark) \kms. The sample spectra are selected to show regions
of multiple spectral features and the central region.
 }
\end{figure*}

{\it The Dispersion Map:} Fig.\,5 displays the dispersion map with a few selected spectra.
The one dimensional dispersion is quite moderate, $\sigma_{\rm 1d} \lapprox 10$ \kms\ in large
parts of the galaxy. Regions of high dispersion are found where the velocity field shows
discontinuities or irregularities, and the sample spectra reveal that here the enhanced velocity
dispersion
is caused by multiple spectral features rather than genuinely broad lines. The velocity shifts 
between the narrow peaks is 150--200 \kms, i.e. the typical rotational velocity of a spiral galaxy. 
There is also a continuous curved region of elevated dispersion (i.e. multiple spectra)
in the center of the map.
Even the width of the spectrum at the intensity peak is quite low (lower panel, right
hand side) with $\sigma_{\rm 1d}=50$ \kms, which is narrow for the center of
a merging system.

{\it Channel Maps:} Dividing the emission into three velocity bins, (2380 -- 2485 \kms),
(2506 -- 2570 \kms), (2590 -- 2720 \kms), (Figs.\,6a,b,c) shows the emission being distributed
in long, arm-like features of quite different appearence in the three bins. The low and high
velocity bins were chosen to include each one of the two features seen in the double peaked
spectra (e.g. middle right panel in Fig.\,5). The middle bin includes emission at intermediate
velocities near the systemic value.

The low velocity map (Fig.\,6a) shows a patchy distribution extending over 20$''$ to the north-south. 
Closer to the center, brighter emission curves 5$''$ in the east-west direction.

At mid-velocities (Fig.\,6b), the distribution is dominated by a striking, elongated curved shape, 
approximately 15$''$ (2.8 kpc) in length. This structure is the main cause of the discrepancy
between the \twco\ emission distribution and the R-band light distribution. It traces a central
prominent dust lane. To show this,  we have overlayed the mid-velocity
\twco\ emission on an archival HST image in Fig.\,7.
Another, fainter, 20$''$ (3.8 kpc) long curved
structure is found in the north. This feature may be a continuation of the central dust lane and
the gap between them a depression in the \twco\ brightness.

\begin{figure*}
\resizebox{16cm}{!}{\includegraphics{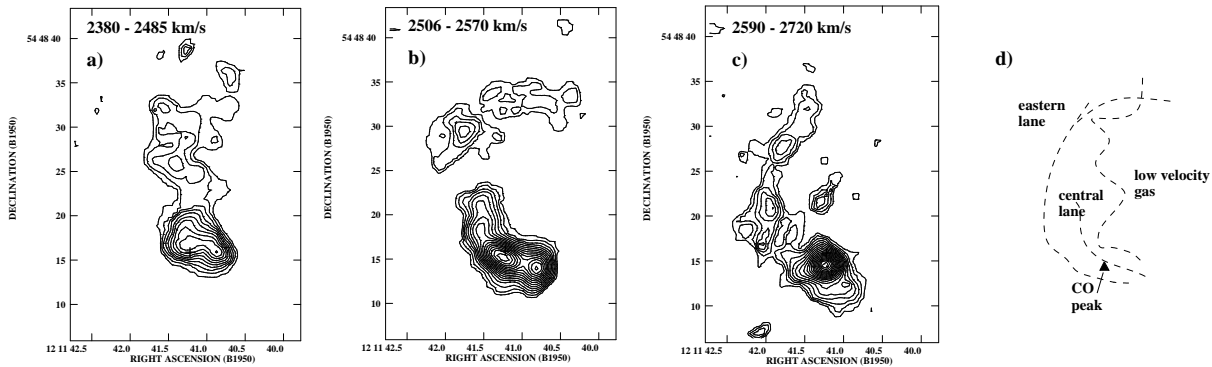}}
\caption{\label{something} The left panel {\bf a)} shows the integrated intensity
(naturally weighted \twco\ map) in the lower part of the velocity range (2380 -- 2485 \kms),
panel {\bf b)} shows the central part of the velocity range (2506 -- 2570 \kms), and
panel {\bf c)} shows the highest velocities (2590 -- 2720 \kms). The dashed lines indicate
the position of the central dust lane from the mid velocity panel.
Contours are (0.7, 1.4, 2.1, 2.8 ... 14) Jy beam$^{-1}$ \kms. The cross marks the
radio continuum position. The right panel {bf d)} is a sketch where the features from the three
velocity bins are indicated with dashed lines.}
\end{figure*}

The high velocity emission bin (Fig.\,6c) shows a bright peak in the center with 
a source size of 2$\hbox{$\,.\!\!^{\prime\prime}$}5 
\times 1\hbox{$\,.\!\!^{\prime\prime}$8}$. A prominent, narrow curved feature stretches out to the north-east, 
(20$''$ to the north) and follows most of the eastern obscuring dust lane seen in Arp's
B-band image.  We suggest the northern, curved feature in the mid-velocity bin is the continuation of
this curve. As mentioned above, the central dust lane may connect to this feature as well. 
In Fig.\,6d
the features in the three velocity bins are marked with dashed lines.

\begin{figure}
\resizebox{5cm}{!}{\includegraphics{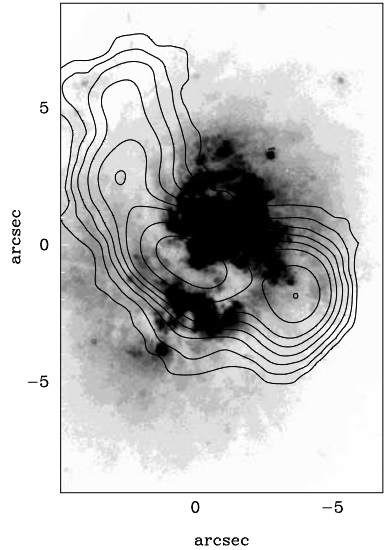}}
\caption{\label{something} Contours of the mid-velocity (2506 -- 2570 \kms) \twco\ emission
overlaid on an archival HST WFPC2 image which shows the \twco\ emission tracing a prominent
central dust lane.}
\end{figure}

\subsubsection{The High Resolution Map --- a close-up on the center}

\begin{figure*}
\resizebox{16cm}{!}{\includegraphics{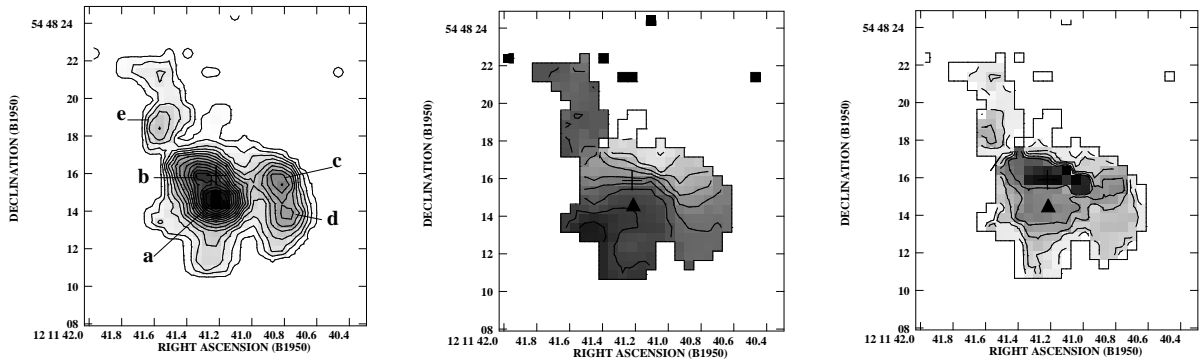}}
\caption{\label{something} {\bf a)} Map of total integrated intensity for the high resolution data
only showing the inner region of NGC~4194. Contours are (0.5, 1.5, 3.5, ...) Jy beam$^{-1}$ \kms. 
The peak flux is 17 Jy beam$^{-1}$ \kms. The intensity peaks are designated $a$ to $e$ with falling
peak intensity. The grayscale ranges from 1 (light) to 16 (dark) Jy beam$^{-1}$ \kms. {\bf b)} The high resolution
velocity field. Contours and grayscale are the same as in Fig.\,2b. The position of the \twco\
peak $a$ is indicated with a black triangle. {\bf c)} High resolution dispersion
map. The position of the \twco\
peak $a$ is indicated with a black triangle.}
\end{figure*}

For a closer look at the structure of emission in the inner 10$''$, we
selected the high resolution data only, and thus increased the resolution to 
$1\hbox{$\,.\!\!^{\prime\prime}$}7$
(but lost some sensitivity and $uv$-coverage).
The integrated intensity map (Fig.\,8a) of the inner region 
consists of 5 peaks, $a$, $b$, $c$, $d$ and $e$ (Table\,3) and extended emission surrounding them.
The total flux of the inner $10'' \times 10''$ is 64 Jy km/s.
The brightest peak, $a$ (17 Jy \kms\ beam$^{-1}$), is at the same position as in the lower
resolution map, i.e.\ close to the radio continuum peak, 
but shifted to the south by $1\hbox{$\,.\!\!^{\prime\prime}$}5$ (280 pc).
Feature $a$ has a size of $1\hbox{$\,.\!\!^{\prime\prime}$}8 \times 
1\hbox{$\,.\!\!^{\prime\prime}$}3$ and a position angle of $45^{\circ}$. The fitted FWHM sizes
of features $a$--$e$ are typically 200--400 pc, which is comparable to the sizes of Giant Molecular
Associations (GMAs) found in other galaxies.
Features $b$ and $c$ can be identified with knots in the H$\alpha$ distribution (Armus \etal\ (1990)). 
The peak intensity in the high resolution map is
200 mJy, which corresponds to 5.4 K.

Fig.\,9 is an overlay of the integrated high resolution \twco\ emission
on an archival HST image (enhanced to show
the dust lane, same as Fig.\,5). The peaks $a$, $b$, $d$ and $e$ all lie on the central dust lane.

The velocity field (Fig.\,8b) is slightly more twisted than in the lower resolution map
and covers the same velocity range. Features $d$ and $c$ are causing twists in the velocity
field because they are at too low velocities to fit a spider diagram. The maximum dispersion (100 \kms) in Fig.\,8c does not occur in $a$, 
but in an elongated (east-west) region located north of the \twco\ peak,
close to feature $b$. This is caused by the crossing of the dust lane, i.e.\ by a superposition
of multiple spectral features.

Fig.\,10 shows an overlay of the \twco\ emission on the 1.49 GHz radio continuum structure
(Condon \etal\ 1990). There is a rough agreement between the CO and the
radio-continuum distribution which seems to 
break down at smaller scales (i.e. less than 1 kpc). The \twco\ peak (feature $a$), for example,
is located $1''$ south of the radio continuum peak. There is an arm-like structure
curving from the center to the west which has a \twco\ counterpart in feature $c$. 
The one \twco\ feature entirely lacking in the continuum distribution is $e$ in
the north.
Comparisons between radio continuum and CO distributions
in starburst galaxies often show a  correlation between the two
distributions (e.g. Yun \etal\ 1994, Bajaja \etal\ 1995). This effect is often attributed
to cosmic ray heating of the molecular gas (e.g. Adler \etal\ 1991; Allen 1992). 
We will discuss this correlation further in Sect.\,4.2.

\begin{figure}
\resizebox{6cm}{!}{\includegraphics{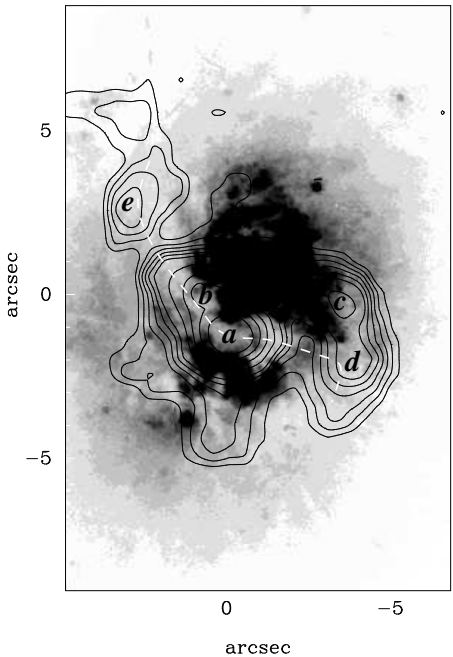}}
\caption{\label{something} Overlay of the integrated high resolution \twco\ emission in contours
on an archival HST image (same as Fig.\,5). The central dust lane is indicated as a white,
dashed line, and most of the \twco\ peaks ($a$, $b$, $d$, $e$) lie on this lane.
The \twco\ peaks are indicated with their corresponding letters.}
\end{figure}

\begin{figure}
\resizebox{7cm}{!}{\includegraphics{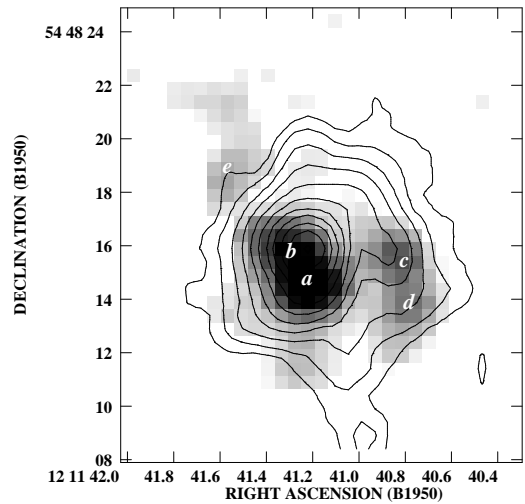}}
\caption{\label{something} Overlay of the radio 1.49 GHz radio continuum (Condon \etal\ 1990) in
contours on the high resolution  \twco\ emission in grayscale. 
The \twco\ features are labeled with their corresponding letter.
The contours are logarthimically spaced, as in Condon \etal. }
\end{figure}

\begin{table*}
\caption{\label{prop} Central Emission Peaks}
\begin{tabular}{lrrcccc}
Peak &
\multicolumn{1}{c}{R.A. (1950.0)} & 
\multicolumn{1}{c}{Dec (1950.0)} &
Peak Intensity  & Size (FWHM) & $V_{\rm c}$ (\kms) & Notes  \\
\hline \\
 & (1950.0) & (1950.0) & (Jy \kms\ beam$^{-1}$) & $''$ & (\kms) & \\
\hline \\
 
$a$ & $12^{\rm h}\ 11^{\rm m}\ 41.22^{\rm s}$ & $54^{\circ}\ 48'\ 14\hbox{$\,.\!\!^{\prime\prime}$}4$ &
17 & $1\hbox{$\,.\!\!^{\prime\prime}$}5$  & 2592 & \twco\ peak \\

$b$ & $12^{\rm h}\ 11^{\rm m}\ 41.28^{\rm s}$ & $54^{\circ}\ 48'\ 15\hbox{$\,.\!\!^{\prime\prime}$}9$ &
13 & $2\hbox{$\,.\!\!^{\prime\prime}$}5 \times 0\hbox{$\,.\!\!^{\prime\prime}$}8$  & 2531 & on dust lane$^{\rm a}$ ; H$\alpha$ knot\\

$c$ & $12^{\rm h}\ 11^{\rm m}\ 40.82^{\rm s}$ & $54^{\circ}\ 48'\ 15\hbox{$\,.\!\!^{\prime\prime}$}4$ &
8.6 & $1\hbox{$\,.\!\!^{\prime\prime}$}7 \times 0\hbox{$\,.\!\!^{\prime\prime}$}$5  & 2496 & H$\alpha$ knot \\

$d$ & $12^{\rm h}\ 11^{\rm m}\ 40.75^{\rm s}$ & $54^{\circ}\ 
48'\ 13$\hbox{$\,.\!\!^{\prime\prime}$}9 &
6.9 & $1\hbox{$\,.\!\!^{\prime\prime}$}8 \times 0\hbox{$\,.\!\!^{\prime\prime}$}7$   & 2539 & on dust lane\\

$e$ & $12^{\rm h}\ 11^{\rm m}\ 41.57^{\rm s}$ & $54^{\circ}\ 
48'\ 18\hbox{$\,.\!\!^{\prime\prime}4$}$ &
4.6 & $2\hbox{$\,.\!\!^{\prime\prime}$}4$  & 2563 & on dust lane \\

\hline 
\end{tabular} \\
a): Double spectrum: peaks at 2464\,\kms\ and 2549\,\kms\ (main) \\
\end{table*}

\subsection{Single dish results}

Results are presented in Table 2 and \twco, \thco\ spectra in Fig.\,12 (HCN was
not detected, and its spectrum is not shown). The \twco\ flux detected by OVRO is, within
calibrational error bars, the same as measured by the OSO 20m telescope, 
which is approximately 60\% more 
than the flux Devereux \etal\ (1994) find with the FCRAO 14m telescope.

We note that the integrated \twco/\thco\
intensity ratio we find is lower, $\approx 20$, than previously reported by Casoli \etal\ (1992),
$\approx 55$, although their error bar of $\pm 22$ includes our measurement. Our result means that
NGC\,4194 is not an extreme HR merger and \thco\ 1--0 is quite detectable, even if the ratio is
clearly above what is usually found for starburst galaxies. There is a hint of
a line shape difference in the \twco\ and \thco\ lines with an extra narrow component at lower
velocities in the \thco\ spectrum. The signal to noise ratio in the \thco\ spectrum is too low to
warrant further speculation on this matter here, but a high resolution \thco\ image may reveal
interesting excitation changes across the galaxy, as has already been shown for the merger
starburst system Arp~299 (Aalto \etal\ 1997; Casoli \etal\ 1999).
The HCN emission is at least as faint as the \thco\ emission with the \twco/HCN intensity ratio
$\gapprox 25$. In comparison to other FIR luminous galaxies (e.g. Solomon \etal\ 1992) the HCN
is relatively faint.

\begin{figure}
\resizebox{7cm}{!}{\includegraphics{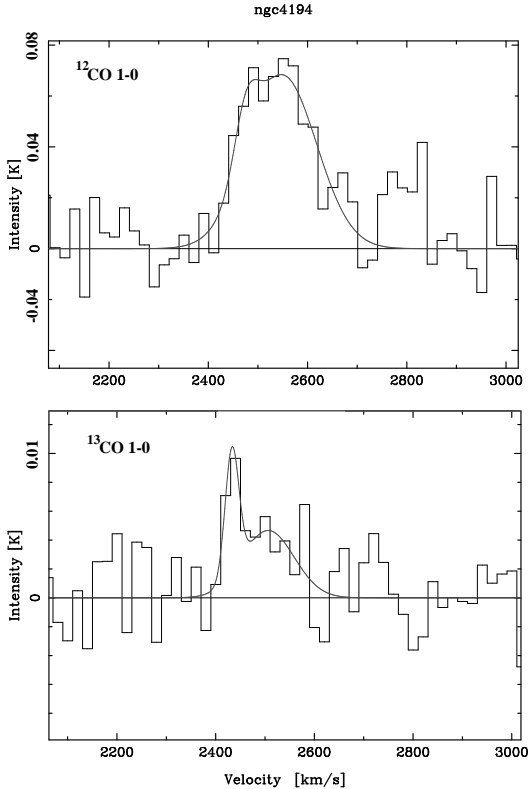}}
\caption{\label{something} The OSO 20m telescope single dish spectrum of \twco\ (upper panel) and
\thco\ (lower panel). The intensity scale is in $T_{\rm mb}$.}
\end{figure}

\section{Discussion}

\subsection{Merger timescale and \twco\ morphology}

\subsubsection{The Medusa as a major disk-disk merger}

Joseph \& Wright (1985) suggested that NGC~4194 is a collision between two disk galaxies and
put it fairly late in a merger
sequence: after Arp~220 and before NGC~3310. They used the apparent diffuseness of the tidal
tail and the coalesced optical main body as principal arguments for this placement 
in the evolutionary scheme. 

However, despite the suggested advanced stage of
the NGC~4194 merger, the molecular gas structure is complex and distributed on a scale of
several kpc. In the late stages of a merger, simulations suggest that the bulk of the molecular gas
would have sunk to the inner 200 pc of the merger, regardless of the previous morphology/orientation
of the precursor galaxies (e.g.\ Barnes \& Hernquist 1996). 
The corresponding size scale for NGC~4194 is an order of magnitude
greater than that. The funneling of molecular gas toward the center of
NGC~4194 has been less effective than for ULIRGs, such as Arp~220 or NGC~6240,
which is interesting since both these mergers appear younger than NGC~4194. Apart from the ULIRGs
being more luminous by factors of at least 10, they also have broader line widths, and deeper
potentials. If NGC~4194 is the result of the merger of two disk galaxies of comparable
mass, i.e.\ a `major merger', either
the merger is less advanced than previously thought, or the intrinsic
and/or orbital properties of the NGC~4194 progenitors are, after all, important factors.

One can use the presence and separation of a double nucleus as an indication of merger age. 
In NGC~4194 there is no obvious sign of another nucleus, but there is
a fairly bright extension to the south of the radio continuum emission, which is the location
of the \twco\ peak, $a$. If $a$ is another nucleus in the merging system, then the
projected separation between them is $1\hbox{$\,.\!\!^{\prime\prime}$}5$, or 280 pc. 
However, the non-compact appearance (and lower lumniosity) of $a$ indicates instead that it is
a bright star forming region rather than a second nucleus, in which case the
upper limit of the nuclear separation becomes even more strict. It is perhaps also possible that
merger orientation hides the second nucleus behind the first one, but this should be a rather
unlikely scenario.

We find it difficult to reconcile the extended, unrelaxed appearance of the molecular gas
distribution with a major merger scenario where the lack of a double nucleus indicates a very
advanced merger. Together with the morphological appearance of the Medusa, this lead us to
suggest another merger scenario:

\subsubsection{The Medusa as a young shell-galaxy}

The southern sharp, curved feature seen in the optical image
may well be a shell in its early stages of formation, thus raising questions on whether
an elliptical/early-type galaxy was one of the progenitors. The total molecular mass
of $2 \times 10^9$ M$_{\odot}$ is indeed less than what one would expect from the collision of
two gas-rich disks.

Quinn (1984) found that shells can form through ``phase-wrapping'' of dynamically
cold disk material in a rigid, spherical potential.  The similarity between the
overall Medusa morphology and simulations of a radial, planar collision between a disk
galaxy and a more massive spherical, is striking (e.g.\ Quinn 1984, Fig.\,1 timestep 6;
Kojima \& Noguchi (KN) 1997, Fig.\,4a timestep 4.7, 6.2). The first shell is forming
on the opposite side of the single, broad tidal tail creating an almost ``cometary''
structure. Although, we note that the implied molecular mass of NGC~4194 suggests that
the disk galaxy involved was more massive than a dwarf --- which is often assumed in
the simulations.

KN also simulated the behaviour of the gaseous component
during the encounter and found that the gas becomes scattered much like the disk stars.
This is in sharp contrast to the models of Weil \& Hernquist (WH)(1993) who instead
conclude that the gas from the disk galaxy
quickly sinks to the center of the merger. This also leads to different
views on the occurence of starbursts in these encounters. In the KN approach, the 
scattering of gas clouds prevents star formation, while WH suggest that
starbursts are likely to occur in the central gas concentration.
In the WH model the shells are expected to be gas-poor because of the separation
between gas and stars. The difference between the KN and WH models is largely caused by
their ISM represenation: KN using sticky particles and WH a hydrodynamical model.

Now, if we adopt the notion that NGC~4194 is indeed a radial collision between a
spiral and a more massive, gas-poor spherical (i.e.\ an unequal mass, or `minor' merger), 
we can make the following observations:
No molecular gas is detected in the southern shell in accordance with the WH model,
but perhaps more sensitive observations can find gas here. Both HI and CO has been detected
in the shells of Centaurus A (Schiminovich \etal\ 1994; Charmandaris, Combes \& van der Hulst
2000).
The molecular gas is not scattered all over the system, nor can all of it be found
in the very inner region. The distribution is somewhere in between the two 
models, and the gas in the inner region seems to be going through a starburst on a fairly
large scale (see Sect.\.4.2). 
The multiple spectral components may reflect gas on different orbits
and in different stages of its dance around the center of the elliptical.
The velocity shift between the two components
is roughly the same as the typical rotational velocity for a spiral galaxy. 
Thus the existence of two 
spectral components does not necessarily imply that the molecular gas is coming from
two different progenitor
galaxies. For both the major and minor merger models, however, 
the presence of two spectral components
suggests that the gas has some time to go before it reaches relaxation.

However, the elliptical-spiral model leaves us with (at least) two quesions: 1) where is the massive
elliptical? 2) what is the significance of the (very) faint tail to the south (which can only
be seen in deep images)?
The first question can better be answered once there is a near infrared K-band image that can
see through the dust lanes and dust enshrouded starburst currently hiding the center.
The southern tail is very diffuse, and may be a remnant of an earlier interaction. Its
significance can be better judged by sensitive HI observations.

We have made a search through galaxy
morphology catalogues looking for Medusa-like mergers. We found three more
objects with
similar morphology (i.e.\ one prominent tidal tail and on the opposite side
of the galaxy a sharp curved feature): ESO 240-G01, NGC~7135 and NGC~5018.
So, the Medusa morphology is not unique, but fairly rare, which could be an effect
of the rather short timescales involved in the radial encounters. 
The Medusa still stands out among these objects because of its bright FIR emission
and hot dust. Perhaps the Medusa is in a short-lived stage of intense star formation
that will consume a large fraction of the molecular gas, leaving the galaxy to look more like
NGC~7135 once the burst is over.

\begin{figure}
\resizebox{7cm}{!}{\includegraphics{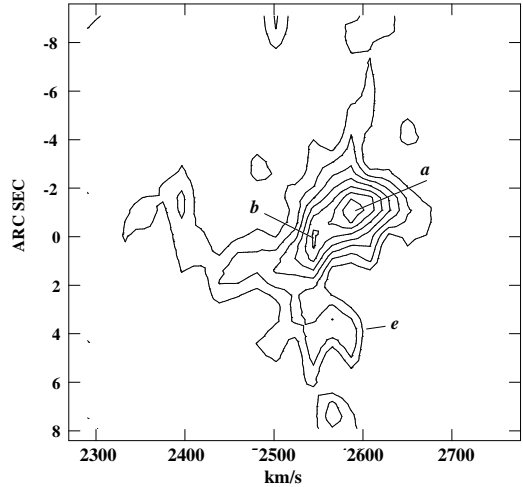}}
\caption{\label{something} position velocity cut along the central dust lane, from $e$ to $a$. }
\end{figure}

\subsection{The starburst}

The starburst of NGC~4194 is extended on a scale of $\approx$ 2 kpc (e.g. Wynn-Williams \& Becklin 1993; 
Armus \etal\ 1990) where also 67\% of the total \twco\ emission can be found.
We suggest that overall solid body rotation of the inner gas (see Fig.\,4) provided a 
dynamical environment favourable to a burst of star formation (Keel 1993).
At least 90\% of the FIR luminosity can be attributed
to a burst of star formation (Prestwich \etal\ 1994; Rowan-Robinson \& Crawford 1989)
and the high $f$(60$\mu$m)/$f$(100$\mu$m) flux ratio (Table 1) indicates that the starburst 
is heating the dust clouds to high temperatures.

Suchkov \etal\ 1993 argue that the correlation between CO and radio continuum distributions
found in galaxies is an effect of cosmic ray heating of the molecular gas. 
They suggest that changing values in the ratio of CO and radio continuum intensity are 
indicators of the starburst age, with young starbursts having a higher value than old bursts
since the cosmic rays are more confined to the molecular layer if the burst is young. 
In this model, the (low energy)
cosmic rays originate in supernove remnants (SNRs). Their electron component is responsible
for the radio continuum emission at cm wavelengths, while their proton component heats the gas. 
This provides a natural explanation for the
rough correlation between continuum and molecular line emission -- both are tied to star
formation activity. This relation is, however, also expected if the (largely non-thermal)
continuum emission simply traces SNRs and UV radiation from massive stars plays the 
major role in the heating of the dust and gas. 

It is of course also possible to find molecular gas without a strong radio continuum 
counterpart if the gas is quiescent and star formation has not (yet) set in. The molecular 
peak $e$ in Fig.\,10 may be an example for such a region.

Bonatto \etal\ (1999) compare NGC~4194 with the starburst merger 
NGC~3256. Their population synthesis results suggest that both galaxies have 
very similar stellar populations.
Starbursts occured both at intermediate (1--2 Gyr) and young ages. 
NGC~3256 also has a kpc scale starburst (Graham \etal\ 1984) and is an HR merger with relatively
faint \thco\ 1--0 emission (e.g. Aalto \etal\ 1991a) and hot dust.

\subsection{The central gas surface densities and the molecular line ratios}

The \twco\ surface brightness of NGC~4194 is lower by an order of magnitude compared to other
HR mergers such as Arp~220 and Mrk~231. The average gas surface density for
the extended starburst region of NGC~4194 is 500 --- 1000 M${_\odot}$ pc$^{-2}$, i.e.\ more than an
order of magnitude below the surface density found in ULIRGs for the same CO--H$_2$ conversion
factor (e.g.\ Bryant \& Scoville 1999).
Addressing the question of the reason for the high ${\cal R}$ posed in the
introduction, we therefore conclude that the turbulent, extreme high pressure scenario
for compact nuclear starburst regions (Aalto \etal\ 1995) is not well suited for explaining 
the unusual molecular line ratio of NGC~4194. Instead, we adopt a similar explanation as for
NGC~3256: high gas temperatures depopulating the lower levels coupled with a disrupted and
fragmented ISM (Aalto \etal\ 1991). One way of testing this scenario is to measure the \twco/\thco\
intensity ratio for higher rotational transitions, and to obtain high resolution images of \twco\
and \thco\ like those for the Arp~299 merger.
We can also conclude that NGC~4194 is the exception to the general trend of increasing central
concentration of the molecular cloud distribution with increasing $f$(60$\mu$m)/$f$(100$\mu$m) flux
ratio (Bryant 1997). A study of the intrinsic properties of compact and kpc scale HR galaxies is
underway (Aalto \etal, in preparation).

The star formation rate (SFR) in NGC~4194 has been estimated to be
$\approx$40 M$_{\odot}$ yr$^{-1}$ (Storchi-Bergmann \etal\ 1994),
based on H$\alpha$ emission. This is not much lower than the estimate
for the HCN-bright ULIRG Arp~220,
$\approx$110 M$_{\odot}$ yr$^{-1}$ for Arp~220 (Smith \etal\ 1998).
Assuming a similar conversion factor for both galaxies, Arp~220 has
about an order of magnitude more molecular gas in its central region.
This would imply a considerably higher star formation efficiency in
NGC~4194. However, the SFR for Arp~220 was determined based on
its FIR luminosity (see Scoville \& Soifer 1991 for the method). Thus,
the estimates are done at different wavelenghts and should
be viewed with some caution. Indeed, if the SFR of NGC~4194
is determined using the FIR luminosity, the result is much lower
(6 -- 7  M$_{\odot}$ yr$^{-1}$), resulting in star formation
efficiencies for NGC~4194 and Arp~220 that are comparable.

The ratio of  L$_{\rm FIR}$/L$_{\rm CO}$ ratio is indeed similar
in both galaxies implying that both starbursts are equally efficient in
producing IR photons per unit molecular mass (again assuming that CO
traces mass in roughly the same way in both objects).

Even if the star formation efficiencies and the  L$_{\rm FIR}$/L$_{\rm
CO}$ are similar, there may be significant differences in the star formation
properties of moderate luminosity mergers and ULIRGs, since the ISM
in the two types of galaxies has a different structure. The starburst of
NGC~4194 takes place in a low pressure environment compared to the
starbursts at the heart of ULIRGs. The lower gas pressure results in a
lower mass fraction of dense clumps and may allow destructive UV
photons to penetrate further into the molecular clouds, where they
can heat the dust and dissociate the molecules.  Then,  the young,
massive
stars in NGC~4194 would have a larger impact on the surrounding
gas, disrupting it more effciently. This may also account
for the relatively high brightness in H$\alpha$ and the hot dust.
NGC~4194 has near solar metallicity, so the apparently  efficient
heating of the dust cannot be explained by a gas-to-dust ratio
that is unusually low (e.g. Bonatto \etal\ 1999;
Storchi-Bergmann \etal\ 1994).

\subsection{The Fate of the Molecular Gas}

The elevated \twco/\thco\ 1--0 molecular line ratio together with the IR properties
indicate that the molecular gas is being strongly affected by the newborn stellar
population. 
The estimated star formation rate of NGC~4194 is about 40 M$_{\odot}$
yr$^{-1}$ which, if it continues, will have consumed the available gas in the center 
within 40 million years.
The unusual molecular cloud properties may well be an indication
that the standard conversion factor is inapplicable, so the real gas consumption time
scale may be lower. The cloud properties will likely be quite similar to those in the
NGC~3256 model (Aalto \etal\ 1991a) where the kinetic gas temperatures are found to be close
to 100 K and the clouds are quite small, a few pc in size. Such a molecular ISM will radiate
more brightly per unit mass than the cool disk clouds for which the conversion factor is calibrated.
Experience from the NGC~3256 model leads us to suggest that the real molecular mass in the
starburst region may be a factor of $\approx$5 lower than the mass based on the `standard'
conversion factor.

If all of the gas is to be consumed, also the gas at higher galactic radii needs to be transported
to the center. The morphology of the central dust lane with respect to the \twco\ peak is very
suggestive of gas transport to the center. A position-velocity cut (Fig.\,12) along the central
dust lane, from $e$ to $a$, shows that the gas on the
dust lane may be moving continuously into the \twco\ peak $a$. On both sides of $a$ along the dust lane
the velocities are lower by 50 \kms\ than in $a$, which could imply that gas is flowing into $a$. 
It is of course also possible that $a$ does not lie on the dust lane, but further to the core
of the galaxy,
in which case this may simply be a projection effect. Then again, the fact that $a$ 
is offset from the
kinematic center argues against this last possibility. There are interesting morphological features
on the dust lane just east and west of $a$ where it seems to flare out into a triangular
shape, and darken (brighten in the negative Figs.). From Fig.\,12 it is also clear that at the
position of $b$ the emission splits into a brighter part at higher velocities and fainter emission
at lower velocities.

The final appearance of NGC~4194 may well be that of an elliptical with shells and a remaining
minor axis dust lane. 

\section{Summary}

We have obtained a high resolution map of the Medusa merger, NGC~4194,
in the \twco\ 1--0 line. The main conclusions we draw from this map and
single dish \twco, \thco\ and HCN observations are as follows:

\begin{enumerate}

\item The molecular distribution in NGC~4194 is extended over almost
      5\,kpc. Thus, this moderate luminosity merger shows clear 
      morphological differences to typical ULIRGs, where an FIR
      luminosity higher by an order of magnitude usually goes along
      with a very compact molecular distribution on a scale of
      $\lapprox$ 1kpc.

\item The structure revealed by our map is complex, consisting of a
      rotating central concentration, low level emission 
      extending north almost into the base of the tidal tail and
      very prominent arcs and filaments that can be identified with
      two dust lanes visible in the optical. 

\item Along the central dust
      lane, we can identify several clumps that may be Giant Molecular
      Associations (GMAs). Some of these GMAs are positionally 
      coincident with radio continuum features. However the strongest 
      molecular concentration visible in \twco\ is offset from  
      the radio continuum peak. This peak may represent a second 
      pre-merger nucleus in the system or -- more likely -- 
      a region of exceptionally strong gas concentration and 
      star formation.

\item NGC~4194 is clearly detected in a single dish \thco\ spectrum.
      Thus, it has a high but not extreme \twco/\thco\ line intensity 
      ratio of $\sim 20$. The extent of the molecular 
      distribution and the relative central surface density of the 
      gas indicate that the physical mechanism causing the high ${\cal R}$
      is different in NGC~4194 from ULIRGs with high values of
      ${\cal R}$. We suggest that a scenario with small clouds having 
      high kinetic temperatures may account for the high ${\cal R}$ found 
      in NGC~4194, but cannot exclude that diffuse gas may also play 
      a role.

\item The non-detection of the high density tracer HCN indicates that
      the amount of dense gas in NGC~4194 is lower than in ULIRGs.
      However, the star formation efficiency may be nearly the same,
      with a gas consumption time of less than 40 million years.
      Alternatively, massive young stars may have a higher impact on 
      the fragmented ISM we suggest for NGC~4194.

\item It is likely that gas is at present flowing along the dust lanes
      toward the nucleus. However, the extended molecular 
      distribution is at odds with the classification of 
      NGC~4194 as an advanced major merger. The two 
      properties can only be reconciled if the gas flow to the 
      center of the system was inefficient compared to ULIRGs. 
      Alternatively, NGC~4194 can be interpreted as an 
      emerging shell galaxy resulting from a merger beween an early type 
      and a spiral galaxy. Comparing simulations of such encounters with 
      the morphology of NGC~4194 lead us to favour this latter interpretation
      of the Medusa merger history.
      
\end{enumerate}

To further our understanding of the merger history of NGC~4194, 
the system needs to be mapped in more molecular transitions as well
as H\,I and H$\alpha$. It represents a class of little studied moderate
luminosity mergers. More generally, knowledge about the parameters
that govern the extent and efficiency of the starburst resulting from
a merger and its connection to the progenitors can only be gained if
the moderate luminosity systems are investigated along with the high 
luminosity ones. If galaxy formation at high $z$ happened through 
mergers, it is essential to understand the differences between 
different types of mergers in the local universe, where the
systems can be resolved with present techniques.

\acknowledgements{Many thanks to Peter Bryant for his help in initiating this study.
We thank J. Mazzarella for generously sharing his R-band image
of NGC~4194 with us. We are thankful for the useful suggestions by an anonymous
referee.
This research has made use of the NASA/IPAC Extragalactic Database
(NED) which is operated by the Jet Propulsion Laboratory, California
Institute of Technology, under contract with the National Aeronautics and
Space Administration. 
The OVRO mm-array is supported in part by NSF grant AST 9314079, and 
by the K.T. and E.L. Norris Foundation.}

\end{document}